# Time of arrival imaging: The proof of concept for a novel medical imaging modality

Tao Feng

*Abstract—* Ionizing radiation causes safety concerns in medical imaging. A transmission scan similar to computerized tomography (CT) but with non-ionizing radiation such as microwave or radio wave can be challenging due to the similarity of the wavelength and object size. It has been shown that with the use of ultra-wideband (UWB) electromagnetic signal and time of arrival (ToA) principle, it is possible to locate medical implants given the permittivity distribution of the body. We propose a new imaging modality using the reverse process to acquire permittivity distributions as a surrogate of human anatomy.

In the proposed systems, the locations of the signal source, receiver, and signal shapes are assumed to be known exactly. The measured data is recorded as the time it takes for the signal to travel from the signal source to the signal receiver. The finite-difference-time-domain (FDTD) method is used for the modeling of signal propagation within the phantom, which is used for both simulation and image reconstruction. Image reconstruction is achieved using linear regression on the training pairs, which includes randomly generated images and its corresponding arrival times generated using the FDTD approach. The linear weights of the training images are generated to minimize the difference between the arrival time of the reconstruction image and the measured arrival time. A simulation study using UWB signal with the central frequency of 300 MHz and the Shepp-Logan phantom was carried out. Ten-picosecond timing resolution is used for the simulation and a multi-resolution approach (from 60 ps to 30 ps to 10 ps) was used for image reconstruction.

The initial simulation study validated that the principle of the proposed approach. The quantitative difference between the arrival times of the phantom and the reconstructed image reduced with an increased iteration number. In the reconstructed images, large structures were resolved in the second resolution setting after 300 iterations. However, the reconstructed image also showed high noise pattern due to the use of randomly generated images as training images. The presence of noise was also gradually reduced with a higher iteration number. The quantitative error of the reconstructed image reached below 10% after 900 iterations, and 8.4% after 1200 iterations. With additional post-smoothing to suppress the introduced noise pattern through reconstruction, 6.5% error was achieved.

In this paper, an approach that utilizes the ToA principle to achieve transmission imaging with radio waves is proposed and validated using a simulation study.

*Index Terms—* Medical imaging, Time of arrival

I. INTRODUCTION

Medical imaging techniques play a vital role in the clinical diagnosis of diseases and facilitate advances in biomedical research. Computerized Tomography (CT) [1], [2] and Magnetic Resonance Imaging (MRI) [3] are among the most widely used anatomical imaging modalities. While CT and MRI have enjoyed tremendous success in related fields, drawbacks have also limited their applications. A CT scan with the use of ionizing radiation has the potential to cause DNA damage, which could increase the risk of cancer [4], [5]. Accidents that lead to radiation poisoning have happened due to human error [6], [7]. An MRI scan, on the other hand, does not impose ionizing radiation to the patient. However, the use of phase encoding prolongs scan duration [8], [9], and the requirement of a strong and uniform magnetic field dramatically increases the cost of MRI scanning. With the presence of strong magnetic fields, strict operation protocols must also be enforced, and serious accidents have occurred due to human error and injuries have been reported [10]. Therefore, a medical imaging modality without ionizing radiation or a strong magnetic field can be a desirable option.

Electromagnetic (EM) radiation is the foundation of most medical imaging modalities except for ultrasound. Based on the energy level, we can classify the EM radiation utilized by current imaging modalities into three categories: low-energy radio frequency waves (Such as in MRI), mid-energy visible light (such as in optical coherence tomography, OCT [11]), and high-energy photons (such as in CT and Positron emission tomography (PET)). The human body is semi-transparent to both low-energy and high-energy EM radiation, making them ideal choices for medical imaging. In transmission imaging such as CT, the high-energy photons behave like particles and travel in straight lines. In contrast, the wavelength of low energy photons is close to the size of human organs. The reflection, refraction, and multipath effects of the low-energy EM radiation invalidate the principle of high-energy transmission tomography. Therefore, in MR imaging which uses low-energy EM radiation, encodings in the K-space is critical for localization.

The use of non-ionizing radiation for medical imaging purpose has been proposed using ultra-wideband (UWB) microwave via space-time beamforming (MIST) [12], [13]. Imaging can be achieved through the measurement of the backscattered signal with the modeling of the instance and scattered electric field. This inverse problem can be ill-posed and often achieved using Born or Rytov approximation [14]. Considerable advances have been made using this technique, and applications in brain stroke detection [15] and breast cancer detection [16] have been proposed.

On the other hand, the combination of UWB signal the radio frequency and the time of arrival (ToA) technique has been proposed for localization of the signal source in the indoor environment [17]. The UWB signal is essentially a signal with narrow full width at half maximum (FWHM) in the time domain. By measuring the time lapse between signal emission and the detection of the first arriving multipath signal, information about the distance between the signal emitter and receiver can be extracted. With the combination of multiple receivers, accurate localization of signal source can be achieved. The same principle has also been applied to the localization of medical implants in human bodies[18], [19]. In the work by Kawasaki et al [19], it has been demonstrated that given the relative permittivity of different body tissues and its distributions (the relative permeability of different body tissues are assumed to be the same), it is possible to determine the location of an implanted medical device. The principle of the proposed medical imaging modality in this paper is similar but the purpose is to reverse the process: the shape of signal and the location of the signal emitter and receiver are assumed to be known exactly, while the distributions of the relative permittivity values ($\varepsilon$) are unknown and to be determined using the ToA principle, combined with multiple acquisitions around the boundaries of the imaging target. Compared with the existing MIST approach above, the use of ToA technique provides another localization approach for transmission imaging using non-ionizing radiations. The ToA technique has already been utilized in imaging modalities such as the time-of-flight (TOF) PET imaging [20]. However, the TOF technique in PET is mainly used for the improvement of sensitivity instead of localization.

TABLE 1

THE RELATIVE PERMITTIVITY OF HUMAN TISSUES

| Tissue | 400 MHz | 900MHz |
|---|---|---|
| Skin | 47 | 41 |
| Fat | 5.6 | 5.5 |
| Muscle | 57 | 55 |

A major theoretical challenge for the proposed approach is to achieve accurate modeling of the propagation of UWB signal inside a complex media (forward model) and its inverse. Even assuming accurate identifications of the first arriving multi-path signal in the receivers, the propagation of the signal won't be along a straight line due to afore-mentioned reasons, therefore the Radon transform commonly used in CT [21] is not applicable. Numerical calculations using the finite-difference-time-domain (FDTD) method [22] have been developed to accurately model the EM propagation in complex media, and the numerical method could be used to replace conventional analytical calculations to obtain the forward model. With the development of machine learning in recent years, it has been shown that the inverse model could be learned from the forward model itself [23], [24], without requiring an explicit inverse model.

In the following sections, we conduct a simulation study in the 2D case and propose an image reconstruction algorithm based on linear regression to recover the distribution of the relative permittivity value from the acquired signal. Acceleration methods of the image reconstruction process are also provided for faster reconstruction. In this study, to demonstrate the proof-of-concept, some practical factors such as the size of the receiver, frequency-dependent permittivity values, and noise in the detected signal are not modeled or considered.

## II. METHOD

*A. Simulation Setups*

In this paper, the FDTD method [22] is implemented based on the software package MIT Electromagnetic Equation Propagation (Meep) [25]. Figure 1 shows the simulation setups, where a Shepp-Logan phantom was simulated. The $\varepsilon$ values of different areas in the phantom can be found in

Table 2. The signal source and receivers are placed at the edge of the phantom. The exact location $(u, v)$ is represented using the equation below:

$$\begin{cases} u = a * \cos(\theta) \\ v = b * \sin(\theta) \end{cases}$$

(1)

where $a$ and $b$ are the two axes of the Shepp-Logan phantom, and $\theta \in [0, 2\pi)$. The size of the two axes of the phantom in this simulation is set to 34.5 cm and 46 cm. A total of 300 discrete values of $\theta$ were simulated for both signal location and receiver location, which makes a total of 300*300 measurements. For easier identifications of the arrival time of the signal at the receiver, the $\varepsilon$ values outside the Shepp-Logan phantom was set to 53 and absorbing boundary condition is used outside the simulation area to avoid misidentification of the light path outside the phantom.

TABLE 2

THE VALUE OF RELATIVE PERMITTIVITY IN THE PHANTOM

| Regions | Relative permittivity |
|---|---|
| R1 | 50 |
| R2 | 16 |
| R3 | 45 |

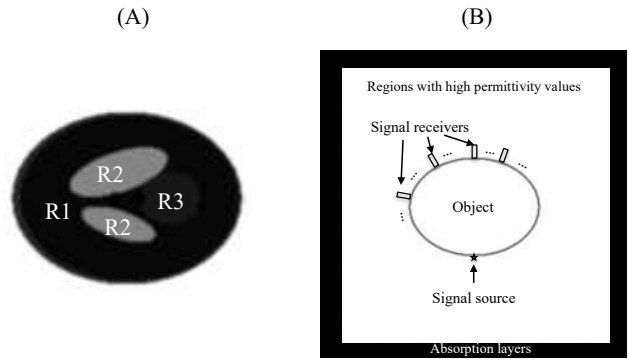

Figure 1 A) Simulated Shepp-Logan phantoms with relative permittivity values can be found in

Table 2.

B. *Simulation setups in this study*

To set up the simulation in Meep, a total simulation area of 1m by 1m was used and the resolution was set to 333 (3 mm spatial resolution). The timing resolution in the simulation was set to 10 ps. The Meep Gaussian source was chosen as the shape of the signal with the central frequency being 300 MHz and width being 450 MHz. The energy of the electric field is used as the detected signal. An example of the original signal and the detected signal from one of the receivers is shown in Figure 2. The envelopes of the signals were obtained by applying a low pass filter to the signal, as shown in Figure 2. In the following sections, the envelopes of the signals are used instead of the original signal unless otherwise specified. In this method, only the arrival time is used for image reconstruction. Therefore, the amplitude of the signal is not used and the conductivity of the media is not reconstructed. As the $\varepsilon$ values are not sensitive at the frequency range used in this study [26], it is assumed to be constant for different frequencies.

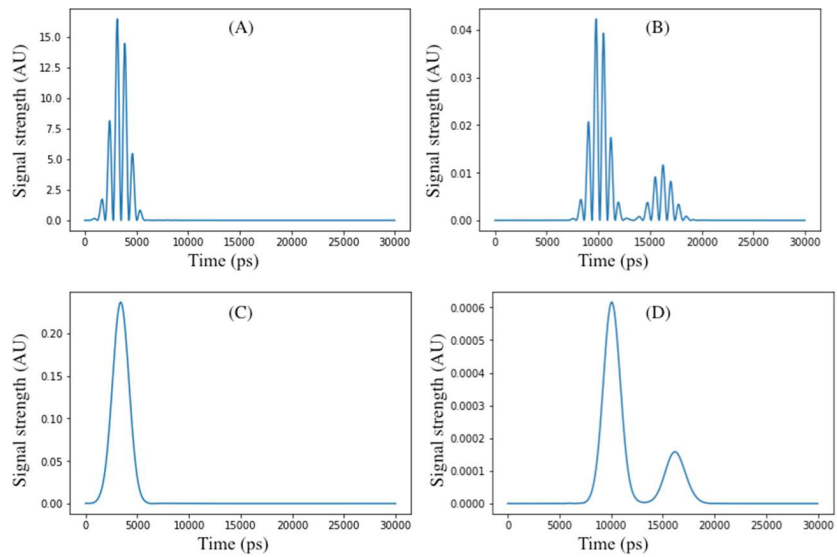

Figure 2 A) the original signal emitted from the source. B) the detected signal by a receiver. C) the envelope of the original signal, D) the envelope of the detected signal by a receiver.

*C. Image Reconstruction*

*B1. Theory*

In a medium with the relative permittivity being $\varepsilon$ and relative permeability, the speed of light is $c/\sqrt{\varepsilon}$. Therefore the arrival time is inversely proportional to the square root of the relative permittivity values, the $\sqrt{\varepsilon}$ image is reconstructed instead of $\varepsilon$.

Let $y$ be the travel time detected at every source-receiver pair and let $x$ be the distribution of $\sqrt{\varepsilon}$. In this paper, for simplicity, the travel time at source-receiver pairs $y$ is noted as the "projection data" and the estimation of the projection data from image $x$ is noted as the "forward projection" process. Unlike CT image reconstruction, the light path between each signal source and receiver is not a straight line due to refractions. Therefore, it is impossible to accurately model the relationship of $y$ and $x$ using a linear equation like the system matrix commonly used in CT and PET imaging. Nonetheless, the "forward projection" process can be represented using the equation below:

$$y = f(x) \qquad (2)$$

While the explicit form of the function $f$ is difficult to derive, the function $f$ can be numerically estimated using the FDTD approach given $x$. The forward projection process $f$ was achieved using the software package Meep.

Image reconstruction process aims to find the image $x'$, whose projection is the same as $y$. In case that $f$ in (2) is an injective function, $x' = x$ can be guaranteed. Let function G represents the inverse of function $f$, so that $x = G(y)$. Similar to (2), the explicit form of $G$ is difficult to derive and it is also challenging to prove or disprove the injectivity of function $f$. Instead, an iterative image reconstruction method with learned image updates is proposed. The image update is acquired using the linear regression approach described below.

Let $\Delta y$ be a small change in the projection data and $\Delta x$ be the corresponding change in the image. With the use of Taylor expansion and only keeping the first linear term, the following approximation can be derived:

$$x + \Delta x = G(y + \Delta y) \approx G(y) + A\Delta y \qquad (3)$$

where $A$ is the Jacobian matrix, with the element $A_{i,j} = \frac{\partial x_i}{\partial y_j}$, where $i$ is the index of the image pixel and $j$ is the index of the projection data. As long as function $f$ is not a linear operation, the values in $A$ are dependent on the projection data $y$ or the image $x$. Therefore $A$ can be written as $A(x)$.

Let $y_0$ be the projection data acquired from the original measurement, $x^{(n)}$ the image acquired using iterative reconstruction methods at n[th] iteration, $y^{(n)}$ its corresponding projection data. Ideally, the projection of the image at the (n+1)th iteration should be $y_0$. To enforce the small change condition in (3), the projection of the image at the (n+1)th iteration is changed to $y^{(n)} + \lambda(y_0 - y^{(n)})$, where $\lambda$ is a sufficiently small number to make (3) approximately true. It can be seen that, as long as the above condition holds, the projection of the reconstructed image will converge to the measured projection. The update image at the (n+1)[th] iteration with the projection equal $y^{(n)} + \lambda(y_0 - y^{(n)})$ of is:

$$x^{(n+1)} = x^{(n)} + \Delta x^{(n+1)} = G\left(y^{(n)} + \lambda(y_0 - y^{(n)})\right) \approx G(y^{(n)}) + A(x^{(n)})\lambda \Delta y^{(n)} = x^{(n)} + A(x^{(n)})\lambda \Delta y^{(n)}$$

(4)

The relationship between the changes in the projection and the changes in the image (matrix $A$) is acquired through a learning processing by utilizing a linear regression model and training samples. Let $x_l$ be a small change of the image $x^{(n)}$, the corresponding change in the projection data $y_l$ satisfies:

$$x^{(n)} + x_l = G(y^{(n)} + y_l) \approx x^{(n)} + A(x^{(n)})y_l$$

(5)

Given $L$ different random images and their resulting projection data, i.e. $L$ training pairs noted as $Y = [y_1, y_2 ..., y_L]$, $X = [x_1, x_2 ..., x_L]$, as long as the values in $X$ and $Y$ are small, the following approximation is valid:

$$X \approx A(x^{(n)})Y$$

(6)

Let $V$ be a vector that minimizes the L2 difference between the "ideal projection update" $\lambda(y_0 - y^{(n)})$ and the "actual projection update" $YV$. $V$ can be calculated by minimizing the difference between the desired update the actual update:

$$V = \underset{V}{\mathrm{argmin}}\left(\lambda(y_0 - y^{(n)}) - YV\right)^T \cdot \left(\lambda(y_0 - y^{(n)}) - YV\right) = \lambda(Y^T Y)^{-1} Y^T \lambda(y_0 - y^{(n)})$$

(7)

Given L random images and their projection data $Y$, $YV$ is the combination closest to the "ideal projection update" $\lambda(y_0 - y^{(n)})$. Therefore, the update equation can be written as:

$$x^{(n+1)} = x^{(n)} + \Delta x^{(n+1)} \approx x^{(n)} + XV$$

(8)

With the use of (8), there is no need to estimate either the function $G$ or the Jacobian matrix $A$. Only the forward projection process is required, which is calculated using the FDTD approach. However, the convergence speed using (8) instead of $G$ or $A$ are expected to be slow due to the small value of $\lambda$ required for this method to work.

The best achievable image resolution using ToA principle as localization method can be roughly estimated using the timing resolution $\tau$. To recover a structure with size $s$ and difference of relative permittivity values being $\Delta\varepsilon$, the straight-line propagation assumption could be used as a rule of thumb, which gives the following equation:

$$\tau c \approx s \left(\frac{1}{\sqrt{\varepsilon}} - \frac{1}{\sqrt{\varepsilon + \Delta\varepsilon}}\right) \approx \frac{s\Delta\varepsilon}{2\varepsilon\sqrt{\varepsilon}}$$

(9)

where c is the speed of light and $\Delta\varepsilon/\varepsilon$ is the contrast of the structure in terms of permittivity. With 10 picoseconds timing resolution, $s\Delta\varepsilon/\varepsilon\sqrt{\varepsilon} \approx 6mm$.

*B2. Generation of Training Pairs*

Equation (8) requires training images with small values $X$ and their corresponding projection data $Y$. The following random images are generated as training images:

i. Images with random values processed by a Gaussian smoothing filter with FWHM randomly chosen between 9 mm and 18 mm (The range is extended to 3 mm to 18 mm after 700 iterations, and further reduced to 1 mm to 6 mm after 900 iterations)

ii. The difference image between the current reconstructed image and the image with a median filter or a Gaussian filter. The size of the median filter is randomly selected from 2 to 15 voxels, and the FWHM of the Gaussian filter is randomly selected from 15 mm to 30mm.

iii. A multiplicative of images generated from scheme i and scheme ii.

The values in the above-generated images are scaled between [-0.3, 0.3] to ensure a small change assumption in (8). The maximum scale is further reduced in large iteration numbers (0.15 for iterations>500 and 0.1 for iterations>700). Approach i is essentially generating Approach ii. is only used after 400 iterations, and approach iii is only used after 700 iterations. The matrix Y is estimated using (2) with Meep. A uniform image (with ε=49) is used as the initial image. The boundaries of the true image are also assumed to be known. No information from the original phantom is used for generating the training images.

*B3. Definition and Acquisition of Projection Data*

The projection data $y$ in this study represents the travel time between a source-receiver pair. Ideally, the location of the first arriving peak can be used to identify the travel time. However, due to the complex media the signal travelled through, possible multi-peak situations created by refractions and reflections are likely to happen. While a simple threshold based approach is shown to be effective in the literature [27], it is not guaranteed to accurately define the first arriving peak. The following scheme is used to calculate each element in $\mathbf{y_0} - \mathbf{y}^{(n)}$ in this study:

i. Use the threshold-based approach to locate the first arriving peak in $y^{(n)}$, let the location be t. The location is determined by fitting a Gaussian function for the signal peak envelop.

ii. Locate the nearest peak in $y_0$, and let the location difference be $\Delta t$

iii. Let the minimal of $\Delta t$ and 1000 ps, and the maximum of $\Delta t$ and -1000 ps represent each element in $y_0 - y^{(n)}$

It can be expected that in case that $x^{(n)}$ is close to the truth, even the first multipath is not selected by step 1, the peaks found in step 2 will approximately match each other in terms of arriving sequence. However, this assumption is not guaranteed. In cases of mismatching arriving sequence, the time difference is likely to be much larger, therefore step 3 is required to reduce contributions from this effect. The upper threshold (1000 ps) is selected based on empirical values.

The residual error E at iteration n is defined as

$$E^{(n)} = \sqrt{(\mathbf{y_0} - \mathbf{y}^{(n)})^T \cdot (\mathbf{y_0} - \mathbf{y}^{(n)})}$$

(10)

The residual error after image update using the randomly generated $X$ is:

$$E'^{(n)} = \sqrt{(\mathbf{y_0} - f(\mathbf{x}^{(n)} + \mathbf{XV}))^T \cdot (\mathbf{y_0} - f(\mathbf{x}^{(n)} + \mathbf{XV}))}$$

(11)

In our update scheme, initially, the training images are a random combination of 8 images generated using the above scheme. As long as $E'^{(n)} > E^{(n)}$, eight more random images and its corresponding projection are generated, and the number of training pairs L is increased by 8. The number 8 was chosen to match the number of threads in the computer used in this simulation and is not expected to affect the outcomes. The update image is only accepted if and only if $E'^{(n)} < E^{(n)}$. A maximum of 200 random images were generated at each iteration.

*B4. Schemes for acceleration*

The use of (8) and iterative reconstruction scheme in the above section can be very slow due to the use of randomly generated matrix $X$. Therefore the following accelerations were employed in this study for faster calculations.

A multi-resolution calculation was used in the FDTD method for the forward projection process in our method. In the first 200 iterations, a spatial resolution of 18 mm and a timing resolution of 60 ps were used. In the following 700 iterations, a spatial resolution of 9 mm and a timing resolution of 30 ps were used. A third resolution setting (3 mm spatial resolution and 10 ps timing resolution) was used after 900 iterations.

The ordered subset approach originally proposed for PET image reconstruction [28] was applied to (7), (10), and (11). In this study, a subset number of 10 was used, i.e. for each iteration, only $1/10^{th}$ of the projection data were used for faster computation. While the use of subset can significantly decrease calculation speed, the use of partial data in (10) and (11) could result in an oscillation of the residual error. Therefore, one subset was used for the estimation of residual errors once oscillation of error was detected.

To reduce the calculation time of matrix $Y$ given $X$, the calculated training pairs from previous iterations with the same subset number were also used together with the new random images.

III. RESULT

Figure 3 shows the distribution of the electric field inside the phantom during simulations. It can be seen that during the first multi-path, the shape of the field is more regular and therefore could be used to extract the distribution of $\varepsilon$. Later signals are more intertwined and much more complicated modelling is required for utilization of this data. Figure 3 confirms the importance to use the ToA principle to isolate the useful signals.

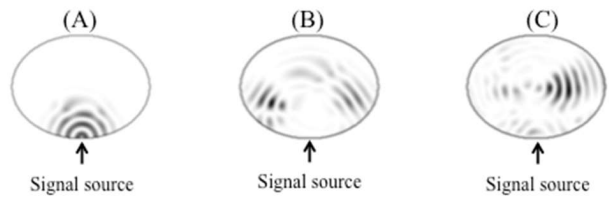

Figure 3 The distribution of the electric field inside the phantom A) moments after signal emission, B) during first multi-path, C) residuals generated by reflections and refractions (non-first-pass signals).

Figure 4 shows the training image generated using the proposed training image generation schemes (i, ii in section B3). Figure 4b) is generated using images at the $400^{th}$ iteration. The training image for scheme i is independent of iteration number. Figure 4A is essentially a noise image, and Figure 4B is the high-frequency component of the current image update.

Figure 5 shows the projection data at the second resolution settings at 300, 500, 700, and 900 iterations. The scale of the projection data was set to be [-1 ns, 1 ns], where black means the arrival time from the simulated phantom data is smaller than the arrival time from the reconstructed image and white represents the inverse. The visual difference between the simulated projection data and the projection of the reconstructed image reduces with an increased iteration number. Even with 900 iteration number, some white/black regions are still present, indicating regions with mismatched peaks of arriving sequence. Figure 6 shows the reconstructed image at 100, 300, 500, 700, 900, 1200 iterations, and 1200 iterations with an additional median filter for noise suppression. With the use of random images in our update scheme, the reconstructed images are quite noisy. However, the general structures of the image become clear after 300$^{th}$ iterations, and the boundaries of the high contrast regions become clear at 900$^{th}$ iteration. The low contrast region R3 in the phantom was not clearly resolved even after 1200 iterations due to the limitation of timing resolution. Based on (9), a timing resolution of 2.5 ps is required to resolve the low-contrast area.

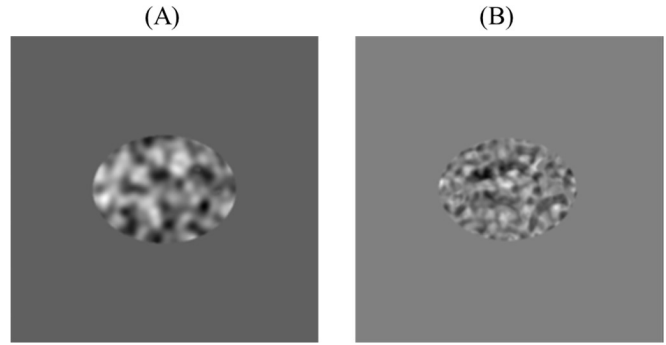

Figure 4 Training image generated using the proposed schemes. a) image generated using random values followed by Gaussian smoothing. b) The difference image between a noisy image and a smoothed image

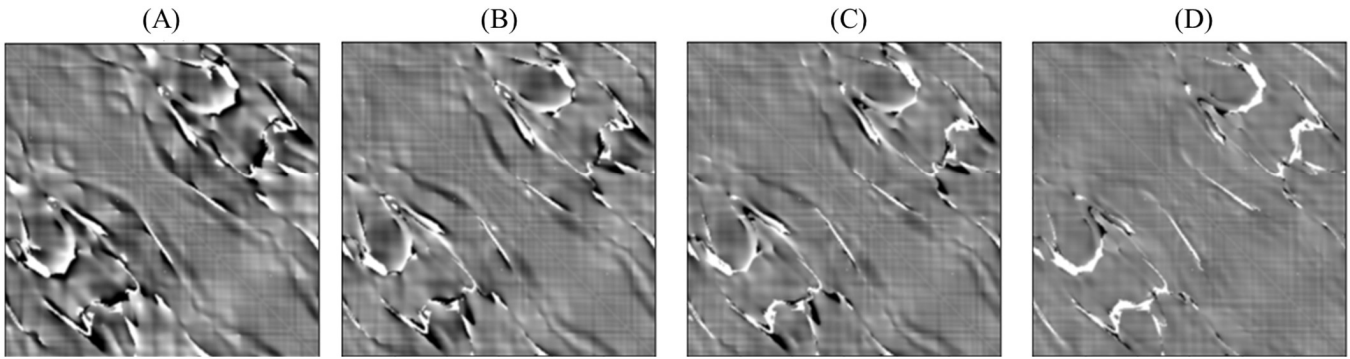

Figure 5 The acquired projection data using the proposed approach at (A) 300, (B) 500, (C) 700 and (D) 900 iteration

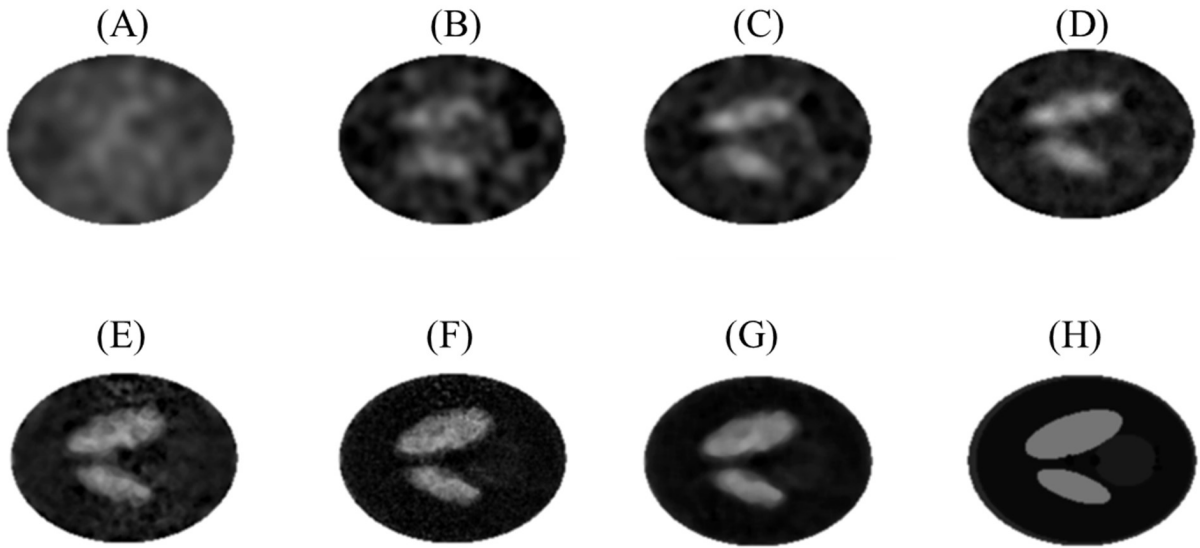

Figure 6 Reconstructed image using the proposed approach at (A) 100, (B) 300, (C) 500, (D) 700, (E) 900, (F) 1200 iteration, (G) 1200 iteration with additional median filtering. The true phantom image is displayed as (H)

The residual error of the projection data using the two resolution levels is also plotted during image reconstruction and is shown in Figure 7. Due to the use of ordered subsets, a fluctuation of residual error with a cycle of 10 iterations is observed. Oscillation of residual error was detected around 400 iterations and one subset was used for calculation of (10) and (11) after 420 iterations. The use of one subset for (10) and (11) was able to enforce a steady decrease in residual error even with the continuous use of 10 subsets for (7). For illustration proposes, the residual error was also multiplied by the subset number. The normalized root mean square error (RMSE) of the reconstructed image is calculated and plotted in Figure 8. With an increased iteration number, the reconstructed image became quantitatively closer to the truth. The quantitative accuracy of the reconstructed image was reduced to below 10% measured by RMSE after 900 iterations (second timing resolution setting). At 1200 iterations, the RMSE of the reconstructed image is reduced to 8.4%(third timing resolution setting). With an additional medial filter (Figure 6(G)) for noise suppression, the RMSE reached 6.5%. The remaining RMSE could be explained by the imperfect timing resolution used in this study. While the residual error was mono-decreasing with the use of 1 subset after 420 iterations, oscillations of RMSE in the images are still observed due to the additional noise introduced through the training images. The general trend of the RMSE is consistent as the residual error. Figure 9 demonstrated an almost linear correlation of residual error in the projection data and normalized RMSE in the images.

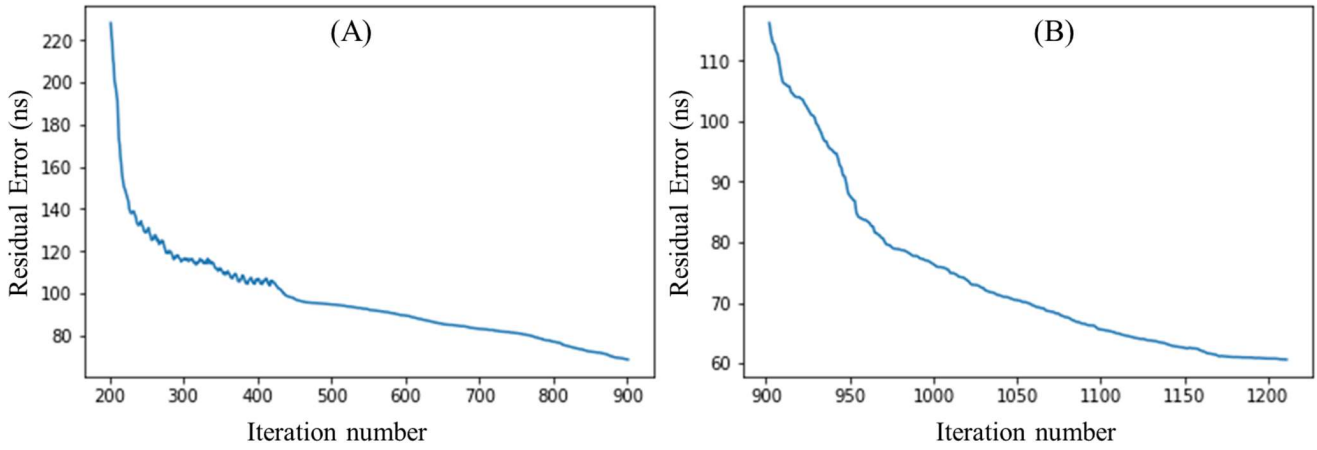

Figure 7 The residual error of the projection data during image reconstruction. A) The second resolution setting. B) The third resolution setting

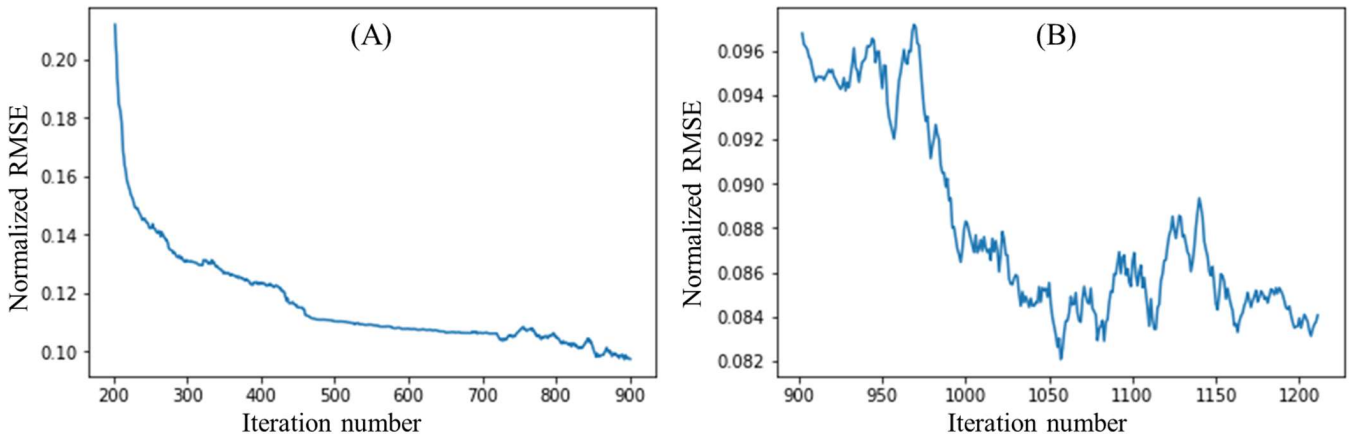

Figure 8 The normalized root mean square error of the image during the reconstruction process. A) The first resolution setting. B) The second resolution setting

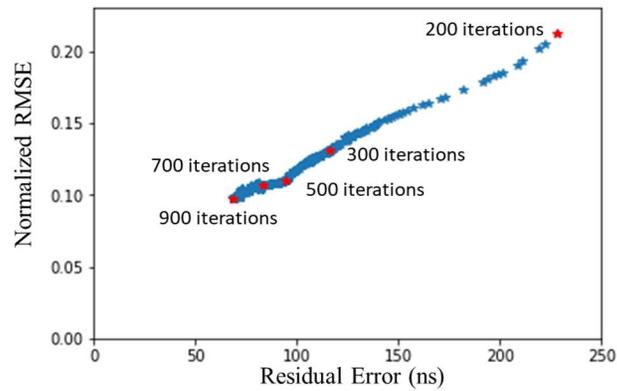

Figure 9 The relationship of residual error in the projection data and normalized RMSE in the image data

## IV. Challenges, Pitfalls, and Limitations

In this simulation study, only the effect of finite timing resolution was modeled. Other practical factors affecting the performance of this study are not modeled and need to be addressed in future studies.

A major practical challenge of our proposed approach is the timing resolution required for this method to work. The results demonstrated while high-contrast structures can be recovered with 10 picoseconds timing resolution, even better timing resolution is required to recover low-contrast regions. The TOF technology has been successfully applied in PET imaging, where the timing resolution of some clinical TOF-PET scanners has been reduced to below 300 picoseconds [20]. Timing resolution around 30 picoseconds has been achieved in experimental stage [29]. Sub-50 picoseconds timing resolution has been reported in near-infrared optical tomography [30] as well. Even better timing resolution can be expected in the recent future with the continuous development of faster electronics, making it possible to resolve finer structures. In this study, the receiver size was also assumed to be zero, which is not realistic. A large receiver size may further impact the resolution of the reconstructed images.

Another challenge is the modeling of the propagation of EM radiations in complex media. In this work, Meep is used for both simulation and image reconstruction. The use of the same software package eliminated possible model mismatch and errors in the FDTD method. The use of fixed $\varepsilon$ values for different frequencies in the signal is also a source of error in the model that is not addressed in this study. The different attenuations effects by the phantom (conductivity) for different frequency components in the signal are also ignored in this study as the absolute amplitude of the detected signal are not used. Nevertheless, the attenuation of the signal could affect the shape of the arriving signal. The effects of model mismatch remain unknown and need to be explored in the future.

The linear approximation in the proposed approach is derived from Taylor expansion, which is only valid with small changes in the projection as well as in the image. The refraction and reflection between layers of different permittivity values could also hinder the detection of the first arriving peak and calculation of measured data. The misidentification of the first arriving peak also invalidates the reconstruction approach, essentially making the function $f$ non-differentiable. While we have shown that the proposed approach works in our simulation settings, a larger permittivity difference in the images could still be problematic. The proposed reconstruction approach also introduced additional noise thanks to the randomly generated training images. more sophisticated regularization approach such as the compressed sensing technique could be used for improved image quality.

The proposed reconstruction approach is similar to the generative adversarial nets (GAN) [31], where the generation of training pairs and the update equation using (8) corresponds to the generative network using noise as input. The peak detection algorithm and the evaluation using the residual error corresponds to the discriminative network. Due to the similarity, it is possible to replace the proposed reconstruction method with a GAN to solve the above-mentioned reconstruction problems.

## V. Discussion and Conclusions

In this study, we demonstrated the proof-of-concept for a new medical imaging modality. This new proposed modality images the relative permittivity values as the surrogate for human anatomy. Neither ionizing radiation nor high magnetic field is required and the theoretical scan speed is limited by the speed of light only. A practical implementation of this imaging procedure could enable cheaper, and safer medical exams.

On the other hand, the challenges and limitations discussed in the above section still need to be addressed for this new modality to be of practical use. In this simulation, with the measured data simulated and reconstructed using 10 ps, larger structures with a moderate contrast of $\varepsilon$ were recovered. It can be expected that with higher timing resolution, smaller structures with low contrast could be resolved.